\documentclass[final,3p,times,twocolumn,fleqn]{elsarticle}

\usepackage{graphicx}
\usepackage{amsmath}
\usepackage{color}

\journal{High Energy Density Physics}

\begin{document}

\begin{frontmatter}





\title{Creation of a homogeneous plasma column by means of\\hohlraum radiation for ion-stopping measurements}
\author[address1]{Steffen Faik\corref{cor1}}
\address[address1]{Goethe-Universit\"at Frankfurt am Main, Max--von--Laue--Str. 1, 60438 Frankfurt am Main, Germany}
\cortext[cor1]{Corresponding author. Tel.: +49 (0)69 798 47846.}
\ead{faik@th.physik.uni-frankfurt.de}
\ead[url]{http://th.physik.uni-frankfurt.de/$\sim$faik/}
\author[address1,address2]{Anna Tauschwitz}
\address[address2]{Helmholtz International Center for FAIR (HIC for FAIR), Max--von--Laue--Str. 1, 60438 Frankfurt am Main, Germany}
\author[address3,address4]{Mikhail M. Basko}
\address[address3]{ExtreMe Matter Institute (EMMI), GSI Helmholtzzentrum f\"ur Schwerionenforschung GmbH, Planckstr. 1, 64291 Darmstadt, Germany}
\address[address4]{Keldysh Institute of Applied Mathematics (KIAM), Miusskaya sq. 4, 125047 Moscow, Russia}
\author[address1,address3]{Joachim A. Maruhn}
\author[address3,address5]{Olga Rosmej}
\address[address5]{GSI Helmholtzzentrum f\"ur Schwerionenforschung GmbH, Planckstr. 1, 64291 Darmstadt, Germany}
\author[address1]{Tim Rienecker}
\author[address4]{Vladimir G. Novikov}
\author[address4]{Alexander S. Grushin}

\begin{abstract}
In this work, we present the results of two-dimensional radiation-hydrodynamics simulations of a hohlraum target whose outgoing radiation is used to produce a homogeneously ionized carbon plasma for ion-beam stopping measurements. The cylindrical hohlraum with gold walls is heated by a frequency-doubled ($\lambda_l=526.5$~$\mu$m) $1.4$~ns long laser pulse with the total energy of $E_l=180~$J. At the laser spot, the peak matter and radiation temperatures of, respectively, $T\approx380$~eV and $T_r\approx120$~eV are observed. X-rays from the hohlraum heat the attached carbon foam with a mean density of $\rho_C=2$~mg/cm$^3$ to a temperature of $T\approx25$~eV. The simulation shows that the carbon ionization degree ($Z\approx3.75$) and its column density stay relatively stable (within variations of about $\pm7$\%) long enough to conduct the ion-stopping measurements. Also, it is found that a special attention should be paid to the shock wave, emerging from the X-ray heated copper support plate, which at later times may significantly distort the carbon column density traversed by the fast ions.
\end{abstract}


%% \pacs{47.40.-x, 64.10.+h, 64.60.My, 64.70.F-}

\end{frontmatter}

\section{Introduction \label{s:intro}}

Today, the heavy-ion stopping in matter at normal conditions \cite{Sig06} is a rather well understood phenomenon. The combination of a high-power Petawatt laser facility and a large-scale accelerator for heavy ions at GSI\footnote{GSI Helmholtzzentrum f{\"u}r Schwerionenforschung GmbH, Planckstr.1, 64291 Darmstadt, Germany, http://www.gsi.de.} offers the unique opportunity to extend this knowledge to dense plasmas at high temperatures. Corresponding experiments with laser-generated plasmas \cite{CoDe.94,RoSt.00,OgOg.01,FrBl.10} are of crucial importance for the indirect drive scenario of heavy ion fusion \cite{HoPl98} and for the ion-driven fast ignition concept \cite{TaHa.94,RoCo.01}, but inevitably the biggest challenge for those experiments is always to avoid spatial non-uniformities in the plasma layer within a sufficiently long lifetime --- on the order of a few nanoseconds --- to measure the stopping power for a bunch of fast ions.

One straightforward way to form a uniformly ionized plasma layer with a constant column density might be to heat a planar foil target --- as usually being used for the measurements with cold matter --- by intense direct laser radiation. Although at a first glance spatial non-uniformities, resulting from the strongly non-uniform intensity distribution across the laser focal spot, seem to be a formidable obstacle, recent theoretical research \cite{TaBa.12} confirmed by experiments \cite{FrBl.13,BoFi.12} has shown that under appropriate conditions sufficiently uniform plasma states, suitable for ion-stopping measurements, can nevertheless be achieved in the direct irradiation scheme.

Another attractive way to generate a uniform plasma state is to heat a sample indirectly with the radiation of a millimeter-scale hohlraum. The hohlraum --- a cavity usually made of a high-Z material, which provides high diffusive resistivity for thermal X-rays --- thereby is directly heated by intense laser pulses to X-ray temperatures of tens and hundreds of electronvolts \cite{LoSi.94,Lin95,Lin98}. The sample may be either placed inside or close to such a hohlraum, usually in the form of a low-density foam. Then the inertial and thermal confinement of the uniformly heated high-temperature plasma can be guaranteed either by the hohlraum itself or by an additional enclosure for a limited period of time.

Theoretical modeling of a hohlraum target is a challenging task for computational physics since it combines multidimensional hydrodynamic simulations with the solution of the spectral transfer equation for thermal radiation. In 2011, 2D (two-dimensional) simulations of two hohlraum configurations, one of them already used in experiments with the NHELIX and PHELIX lasers at GSI \cite{ScBl.11}, were reported \cite{BaMa.11}. In the present work, we present the results of new 2D simulations of another hohlraum target which is representative for undergoing experiments with the PHELIX laser and the UNILAC ion accelerator at GSI \cite{RoBa.11}. The configuration consists of a simple cylindrical hohlraum with gold walls and an empty interior. One hole of the hohlraum serves as the laser beam entrance. At the other hole, a low-density CHO foam inside a copper holder is attached to measure the ion stopping in a partially ionized carbon plasma at moderate temperatures of $T\approx20-30$~eV. A similar setup was already used in earlier experiments at the OMEGA laser facility (LLE, Rochester) to perform X-ray scattering measurements of the heating and cooling dynamics of a carbon foam at higher plasma densities and temperatures \cite{GrGl.08}.

This paper is organized as follows. In Section \ref{s:ralef} the employed radiation-hydrodynamics code RALEF-2D together with the equation of state and spectral opacities are briefly described. Section \ref{s:setup} gives an overview of the target geometry used in the experiments and of the corresponding numerical setup for the simulations, which have been performed in two steps. In the first step, in Section \ref{s:emptyhr}, the heating and radiation of the empty hohlraum without the foam sample are analyzed. In the second step, in Section \ref{s:foamhr}, the plasma evolution of the foam sample is studied and the implications for the ion-stopping experiment are discussed.

\section{RALEF-2D \label{s:ralef}}

\subsection{Two-dimensional radiation hydrodynamics}

All presented results have been obtained with a newly developed radiation-hydrodynamics code RALEF-2D (Radiative Arbitrary Lagrangian-Eulerian Fluid dynamics in two Dimensions) \cite{BaMa.10}, whose hydrodynamics part is based on an updated version of the CAVEAT hydrodynamics package \cite{CAVEAT}. The one-fluid one-temperature hydrodynamic equations are solved in two spatial dimensions (in either Cartesian $(x, y)$ or axisymmetric $(r, z)$ coordinates) on a multi-block structured quadrilateral grid by a second-order Godunov-type numerical scheme. Mesh rezoning and remapping is performed within the Arbitrary Lagrangian-Eulerian (ALE) approach to numerical hydrodynamics.

Thermal conduction and radiation transport have been implemented by newly developed algorithms within the unified symmetric semi-implicit approach \cite{LiGl85} with respect to time discretization. For the thermal conduction, a conservative, second-order accurate symmetric scheme on a 9-point stencil \cite{BaMa.09} is used. For the radiation energy transport, the quasi-static transfer equation
\begin{equation}\label{ralef:transfer-eq}
  \Omega\cdot\nabla I_{\nu}=k_{\nu}\left(B_{\nu}-I_{\nu}\right)
\end{equation}
for the spectral radiation intensity $I_{\nu}=I_{\nu}\left(t,\boldsymbol{x},\boldsymbol{\Omega}\right)$ is solved numerically in order to couple the radiative heating term
\begin{equation}\label{ralef:rad-heating}
  Q_r=-\nabla \cdot \int_0^\infty{d\nu}\int_{4\pi}{I_{\nu} \boldsymbol{\Omega}\,d\boldsymbol{\Omega}}
\end{equation}
to the hydrodynamic energy equation. Within the quasi-static approximation the term $c^{-1}\partial I_{\nu}/\partial t$ (where $c$ is the speed of light) is omitted in equation (\ref{ralef:transfer-eq}). Spatial discretization and integration of equation (\ref{ralef:transfer-eq}) is achieved by the classical $S_n$ method \cite{Car63} to treat the angular dependence of the radiation intensity $I_{\nu}\left(t,\boldsymbol{x},\boldsymbol{\Omega}\right)$, and by the method of short characteristics \cite{DeVo02}, which ensures that every grid cell automatically receives the same number of light rays. The correct reproduction of the diffusion limit on distorted non-orthogonal grids \cite{LaMo.87} is guaranteed by a special combination of the first- and second-order interpolation schemes in the finite-difference approximations to equations (\ref{ralef:transfer-eq}) and (\ref{ralef:rad-heating}).

Energy deposition by a monochromatic laser beam is described by means of the inverse bremsstrahlung absorption. Numerically, propagation of the laser light is treated within the same algorithm as the radiation energy transport --- which means without refraction.

\subsection{Equation of state and opacities}

The equation of state, thermal conductivity, and spectral opacities used in the present work were provided by the THERMOS code \cite{NiNo.05}, which has been developed at the Keldysh Institute of Applied Mathematics (Moscow). The spectral opacities are generated by solving the Hartree-Fock-Slater equations for plasma ions under the assumption of equilibrium level population. In combination with the equilibrium Planckian intensity $B_{\nu}$, used in (\ref{ralef:transfer-eq}) as the source function, the latter means that the radiation transport is treated in the local thermodynamic equilibrium (LTE) approximation. The applicability of the LTE approach to our problem is discussed below in section~\ref{s:emptyhr}.

\begin{figure}
\centering
\includegraphics*[width=0.9\columnwidth]{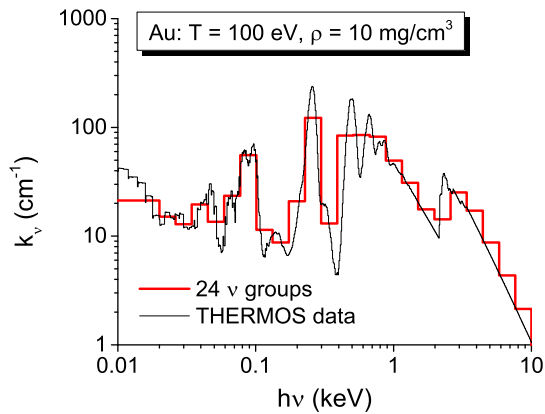}
\caption{\label{f:knu_Au} Spectral absorption coefficient $k_{\nu}$ of gold at $T=100$~eV and $\rho=10$~mg~cm$^{-3}$ used in the simulations: the original \mbox{THERMOS} code data (thin solid curve) are shown together with the group-averaged values for 24 (thick solid curve) selected spectral groups.}
\end{figure}

The transfer equation (\ref{ralef:transfer-eq}) is solved numerically for a selected number of discrete spectral groups $\left[\nu_j,\nu_{j+1}\right]$, with the original THERMOS absorption coefficients $k_\nu$ averaged inside each group $j$ by using the Planckian weight function. Two different sets of frequency groups are prepared for each code run: the primary set with a smaller number of groups (24 in the present simulations) is used at every time step in a joint loop with the hydrodynamic module, while the secondary (diagnostics) set with a larger number of groups (300 in the present simulations) is used in the post-processor regime at selected times to generate the spectral output data. Figure~\ref{f:knu_Au} gives an example of the spectral dependence of $k_\nu$ for a gold plasma at $T=100$~eV and $\rho=0.01$~g~cm$^{-3}$, together with the 24 group-averaged values.

\section{Target geometry and simulation setup \label{s:setup}}

\subsection{Three-dimensional experimental target geometry}

\begin{figure}
\centering
\includegraphics*[width=0.9\columnwidth]{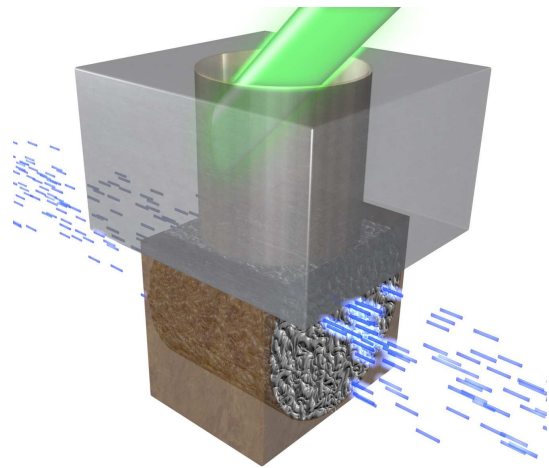}
\caption{\label{f:3D-setup} (color online) 3D scheme of the hohlraum-foam target: shown is a cylindrical gold hohlraum embedded into an aluminum block; the PHELIX laser (green) shoots from the top; a cylindrical cellulose-triacetate (C$_{12}$H$_{16}$O$_8$) foam at the bottom of the hohlraum is surrounded by a copper holder; after creation of the hot plasma, the stopping power for the ion beam (blue dashes) is measured.}
\end{figure}

\begin{figure}
\centering
\includegraphics*[width=0.9\columnwidth]{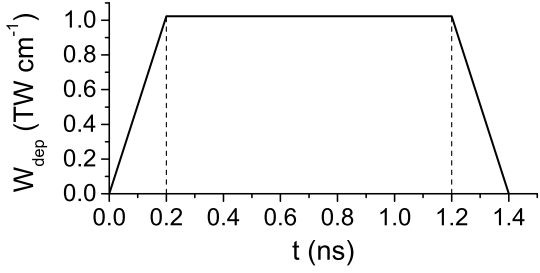}
\caption{\label{f:PHELIX-pulse} Temporal profile of the deposited laser power $W_{dep}$ for the simulated two-dimensional configuration.}
\end{figure}

A three-dimensional (3D) configuration of the combined hohlraum-foam target which covers all of the essential physical processes involved in the experimentally used targets \cite{RoBa.11} is shown in Fig.~\ref{f:3D-setup}. The hohlraum wall is a thin cylindrical gold layer with the inner diameter \mbox{$d_{hr}=1.3$~mm} and the length $l_{hr}=2.0$~mm, embedded inside a massive aluminum block. A $1.4$-ns long frequency-doubled ($\lambda_l=526.5$~$\mu$m) PHELIX laser pulse with $0.2$~ns long flanks (Figure \ref{f:PHELIX-pulse}) shoots at an angle of $\alpha_l=45^{\circ}$ with the total pulse energy of $E_l=180$~J to the center of the hohlraum wall.  In the simulations the spatial profile of the laser intensity was approximated by a Gaussian curve with a full width at half maximum of $0.2$~mm.

At the bottom of the hohlraum, a cylindrical sample made of a cellulose-triacetate (C$_{12}$H$_{16}$O$_8$) foam with initial mean density $\langle\rho_f\rangle=2.0$~mg/cm$^3$, diameter \mbox{$d_f=1.5$~mm}, and length $l_f=d_{hr}=1.3$~mm is attached. Except for the entrance and exit holes for the ion beam (blue dashes in the figure) and the connecting hole to the hohlraum, the foam is surrounded by a solid copper holder. The chosen foam density is, on the one hand, sufficiently low to ensure efficient propagation of the hohlraum radiation, and, on the other hand, high enough to allow fabrication of stable 3D samples by temperatures up to $220$~$^{\circ}$C; the finest pore structure within the foams cellular network is on the order of $1$~$\mu$m. Note, that the column density of the used foam configuration corresponds to that of a $1.3$~$\mu$m thick solid carbon foil, but the combination of the low density and the large thickness of the foam provides the advantage of a diminished hydrodynamic expansion on the nanosecond time scale in comparison to the foil.

In the experiments, the ion bunches of duration \mbox{$t_b=3$~ns} and diameter $d_b=0.5$~mm probed the hot plasma, and the heavy-ion energy losses in the ionized sample measured by the time-of-flight method were compared with the previous ion shot through the cold foam. Simple evaluation of the corresponding Coulomb logarithms indicates that, for the $4.77$-MeV/u titanium ions used in the experiments, the stopping power of the hot foam (approximated as 3.75-times ionized pure carbon at $\rho =2.0$~mg/cm$^3$ and $T\approx 25$~eV) should exceed the corresponding cold value by some 75\%.

\subsection{Two-dimensional simulation setup}

\begin{figure}
\centering
\includegraphics*[width=0.9\columnwidth]{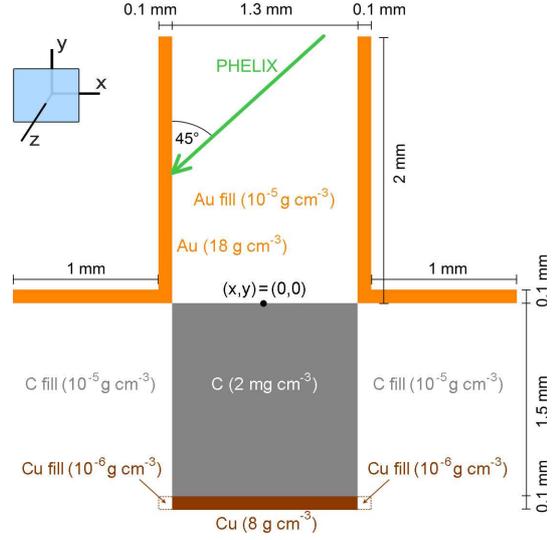}
\caption{\label{f:2D-setup} (color online) Lateral cut of the simulated 2D configuration; all dimensions and initial densities are given; the whole configuration extends to infinity along the $z$-axis; the hohlraum is represented by two gold plates, the foam by a carbon block supported by a copper plate; for simplicity two horizontal gold (rather than aluminum) plates are used to ensure the upper confinement of the lateral plasma expansion.}
\end{figure}

Theoretical modeling of the above described target with an oblique (with respect to the hohlraum axis) incidence of the laser beam is intrinsically a 3D problem. Having no 3D radiation-hydrodynamics code at hand, we had to reduce it to two dimensions. Figure~\ref{f:2D-setup} shows the corresponding lateral cut of the actually simulated 2D target.

Numerical simulations were performed in the $(x,y)$ Cartesian coordinates, where the 3D extension of the simulated region spreads to infinity along the $z$-axis. The 2D hohlraum wall is represented by two gold plates, each with the initial density $\rho_{Au}=18$~g/cm$^3$. The cellulose-triacetate foam was modeled as a homogeneous block of pure carbon with the initial foam density of $\rho_{C}=2$~mg/cm$^3$, supported from below by a copper plate with $\rho_{Cu}=8$~g/cm$^3$. Two additional horizontal gold plates are added to confine the lateral plasma expansion. All the dimensions indicated in Fig.~\ref{f:2D-setup} coincide with the 3D experimental values. The supposedly empty parts of the simulated domain were initially filled with corresponding gases at sufficiently low densities, so that their overall dynamic and thermal influence were negligible. Simulations started from the initial state of pressure equilibrium among all the target parts; the boundary condition of free outflow was applied at all the outer edges.

Spectral energy transport by thermal radiation was treated with 24 logarithmically equidistant (except for the first group) discrete frequency groups delimited by the photon energies
\begin{equation}
  h\nu_j=10^{-4},~0.02,~0.026,~...,~10.0~\textrm{keV},
\end{equation}
as is shown in Fig.~\ref{f:knu_Au}. The delimiting frequencies for the 300 diagnostics groups were logarithmically uniformly distributed between $0.01$~keV and $10$~keV. The angular dependence of the radiation intensity was calculated with the $S_{30}$ method, which offers 960 discrete ray directions over the entire $4\pi$ solid angle.

\subsection{Rescaling of the laser input energy from 3D to 2D}

In 2D simulations we used the same spatial and temporal laser power profiles as in the experiments, and only the total input energy was rescaled to ensure that in two dimensions the hohlraum walls absorb approximately the same amount of energy per unit surface area and exhibit similar dynamics of wall evaporation as in the original 3D configuration. For the 3D input energy of  $E_l=180$~J our rescaling procedure, described in \ref{s:2D3Dconv}, yields (after one iteration) a 2D input energy of $\tilde{E}_l^{(1)}=122.8$~J/mm, which was used in all 2D simulations. The 2D runs needed to iterate $\tilde{E}_l$ were performed for the isolated hohlraum without the attached carbon block. The corresponding areas of the hohlraum walls and holes, entering the equations (\ref{3D2D:Escal}) and (\ref{3D2D:qfirstorder}), are
\begin{eqnarray} \label{S_w=}  &&
  S_w=\pi d_{hr}l_{hr}, \qquad S_h=\frac{1}{2}\pi d_{hr}^2,
  \\ \label{til_S_w=} &&
  \tilde{S}_w=2l_{hr}, \qquad \tilde{S}_h=2d_{hr}.
\end{eqnarray}

\begin{figure*}
\centering
\includegraphics*[width=0.9\textwidth]{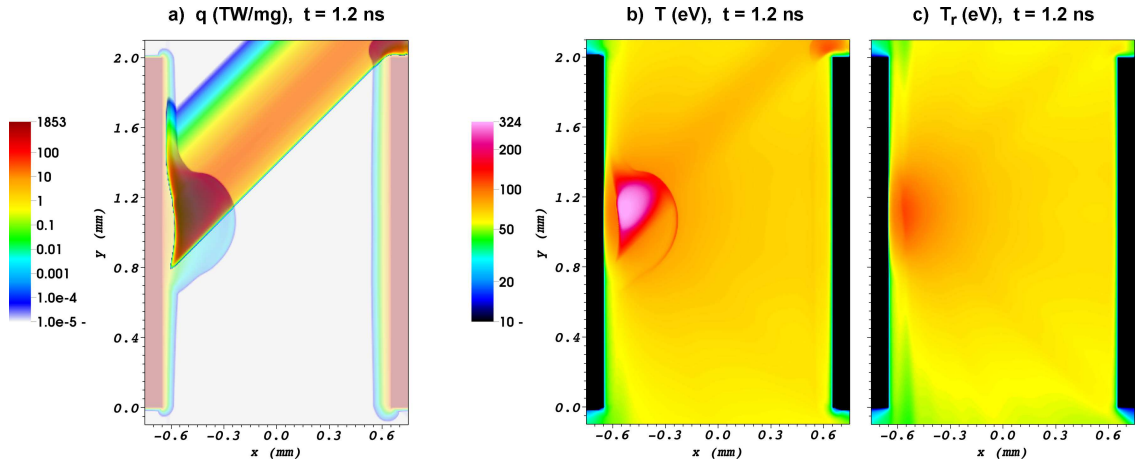}
\caption{\label{f:las-spot} (color online) Color contour plots of a)~the deposited laser power $q$ (semi-transparent plot superimposed on the density), b)~the matter temperature $T$, and c)~the radiation temperature $T_r$ inside the hohlraum by the end of the laser pulse at $t=1.2$~ns; by this time the expanding hot plasma plume at the laser spot on the left wall is heated up to the maximum temperatures $T_{max}=324$~eV and $T_{r,max}=114$~eV.}
\end{figure*}

In Section \ref{s:emptyhr} we present the results for the isolated hohlraum obtained with the first-order input energy $\tilde{E}_l^{(1)}$ on a rectangular mesh with $\simeq 265000$ cells in the purely Eulerian mode. The simulation which included the carbon plasma was performed in the ALE mode with Lagrangian interfaces at the material boundaries on a mesh of $\simeq 110000$ cells and is discussed in Section \ref{s:foamhr}.

\section{Radiative properties of the isolated hohlraum \label{s:emptyhr}}

The heating of the carbon plasma is determined by the thermal history of the hohlraum, especially during the laser pulse. Figure~\ref{f:las-spot}a demonstrates how the laser light is absorbed near the end of the laser pulse, at $t=1.2$~ns; Figs. \ref{f:las-spot}b and \ref{f:las-spot}c show the matter and radiation temperatures inside the hohlraum at this moment. A plume of hot laser-ablated gold plasma rises from the left wall with an average velocity of $\approx 3.5 \times 10^7$~cm/s. Also, some plasma is evaporated from the upper corner of the right wall ``licked'' by the periphery of the laser beam. In the simulated case the net shielding effect of this edge plasma was negligible: it blocked less than 1\% of the total laser pulse energy. 

It should be noted here that actual experiments were performed with different beam-hohlraum configurations --- including those where the laser beam entered the hohlraum horizontally through a hole in the side wall --- and always a special care was taken for the edge plasma not to hinder penetration of the laser light into the hohlraum. The focus position was adjusted such that the laser beam had negligible divergence along the last $\simeq 1$~mm of its path (In the experiment a lens was used with a focal length of $4$~m and a Rayleigh length above $1$~mm.); accordingly, the simulated beam was always assumed to be purely cylindrical without any spatial divergence.

The temperature inside the expanding gold plasma cloud reaches its peak value $T\approx 0.38$~keV at $t=0.2$~ns, the density varies in the range \mbox{$\rho\approx 2\times10^{-4} - 5\times 10^{-3}$} g/cm$^3$, and the LTE ionization degree amounts to \mbox{$Z\approx 45-55$}, which implies a free electron density in the range of $n_e \approx 10^{19} - 10^{21}$~cm$^{-3}$. Recall that the critical electron density for a $\lambda= 526.5$~nm laser light is \mbox{$n_{e,cr} \approx 4 \times 10^{21}$~cm$^{-3}$}. For $Z = 40$, the critical density of free electrons in a gold plasma corresponds to the mass density of $\rho_{cr} = 3.3 \times 10^{-2}$~g/cm$^3$.

\begin{figure}
\centering
\includegraphics*[width=0.89\columnwidth]{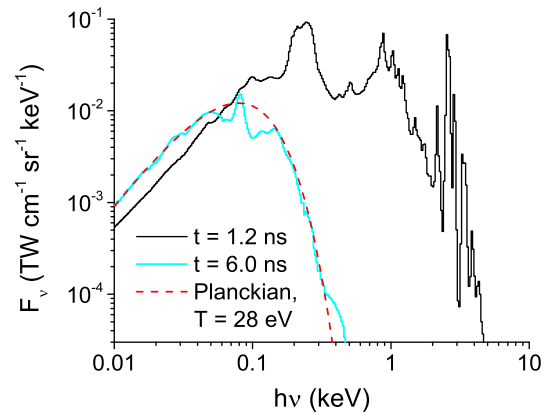}
\caption{\label{f:hr-spectrum} Calculated X-ray spectrum as would have been observed through the lower hohlraum hole and as seen by the foam sample at $t=1.2$ and $6$~ns; the dashed line shows a Planckian fit for $T=28$~eV.}
\end{figure}

\begin{figure}
\centering
\includegraphics*[width=0.89\columnwidth]{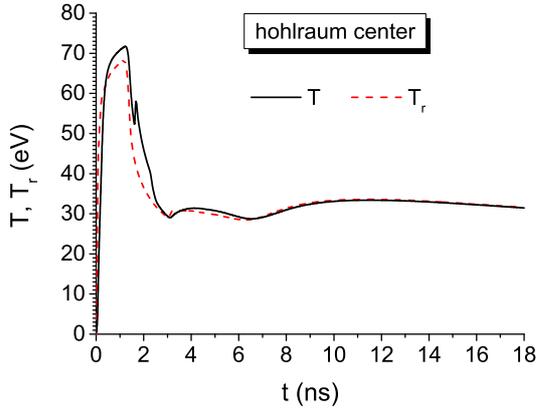}
\caption{\label{f:hr-T-his} Temporal evolution of the average matter temperature $T$ and radiation temperature $T_r$ at the center of the hohlraum (within a box of size $0.2\times0.2$~mm) on a time scale of 18~ns.}
\end{figure}

The calculated X-ray spectra, as would have been observed through the lower hohlraum hole at $t=1.2$ and $6$~ns, are shown together with a Planckian fit of the second spectrum for $T=28$~eV in Fig.~\ref{f:hr-spectrum}. This figure displays the spectral power $F_\nu$ [TW cm$^{-1}$ sr$^{-1}$ keV$^{-1}$] per unit cylinder length, obtained by integrating the spectral intensity $I_\nu$ along an imaginary observation slit (perpendicular to the y-axis) over the range \mbox{$-0.65$~mm~$<x<+0.65$~mm}. Taking its origin from the groups of strong emission lines of gold (cf.\ Fig.~\ref{f:knu_Au}), the highly non-Planckian spectrum at $t=1.2$~ns exhibits three peaks between $0.18-0.3$~keV, $0.78-0.95$~keV, and $2.5-2.7$~keV. As will be shown in the next section, such a spectrum is quite favorable for the efficient supersonic and quasi-volumetric heating of the carbon foam.

On a longer time scale, displayed in Fig.~\ref{f:hr-T-his}, the matter and radiation temperatures inside the hohlraum come close to equilibrium shortly after the laser is off, and for $t \gtrsim 3$~ns stabilize at $T\approx T_r \approx 30$~eV: at this stage the hohlraum gradually cools down due to the radiative energy loss and the calculated X-ray spectrum at $t=6$~ns approaches the Planckian shape. The average matter and radiation temperatures near the hohlraum center, plotted in Fig.~\ref{f:hr-T-his} for times $t\leq18$~ns, were calculated by averaging over the grid-cell masses in the central region (\mbox{$-0.1$~mm~$<x<0.1$~mm}, \mbox{$0.9$~mm~$<y<1.1$~mm}).

These results allow us to assess the applicability of the LTE model to our problem. In general, non-LTE effects may only become significant when (and where) the local radiation spectrum strongly deviates from that of a black-body with the local matter temperature $T$. In our case this happens only during the short period $t\leq 1.4$~ns of laser illumination and only in the vicinity of the laser spot. As a consequence, the non-LTE physics can be expected to affect the ionization equilibrium of gold in the laser focal spot and to modify the spectral details (especially the line features at $h\nu \gtrsim 2$~keV) of the primary hard X-ray flash, generated at $t\leq 1.4$~ns. However, once we focus our attention on later times $t>2$~ns  when the radiation field everywhere (including the carbon foam discussed below) comes close to local equilibrium with matter (see Figs.~\ref{f:hr-spectrum} and \ref{f:hr-T-his}), the use of the LTE approximation is fully justified.

At $t\approx7$~ns, the central region of the hohlraum is slightly reheated to $T\approx T_r\approx33$~eV due to the collision of the expanding clouds of the ablated material from the two hohlraum walls. Such a collision leads to the formation of a strongly radiating shock front \cite{ZeRa02} with a practically full conversion of the kinetic energy into thermal radiation, accompanied by a strong plasma compression. As a result, a thin and dense filament of shock-compressed gold plasma --- clearly visible in Figs.~\ref{f:foam-VisIt}a and \ref{f:foam-VisIt}b below --- is formed. In our case this filament stays close to the hohlraum center, which agrees well with the experimental X-ray pinhole images showing a bright spot at the hohlraums center \cite[Fig.~4b]{RoBa.11}.

\section{Dynamics of the carbon plasma \label{s:foamhr}}

\subsection{Heating dynamics}

\begin{figure}
\centering
\includegraphics*[width=0.9\columnwidth]{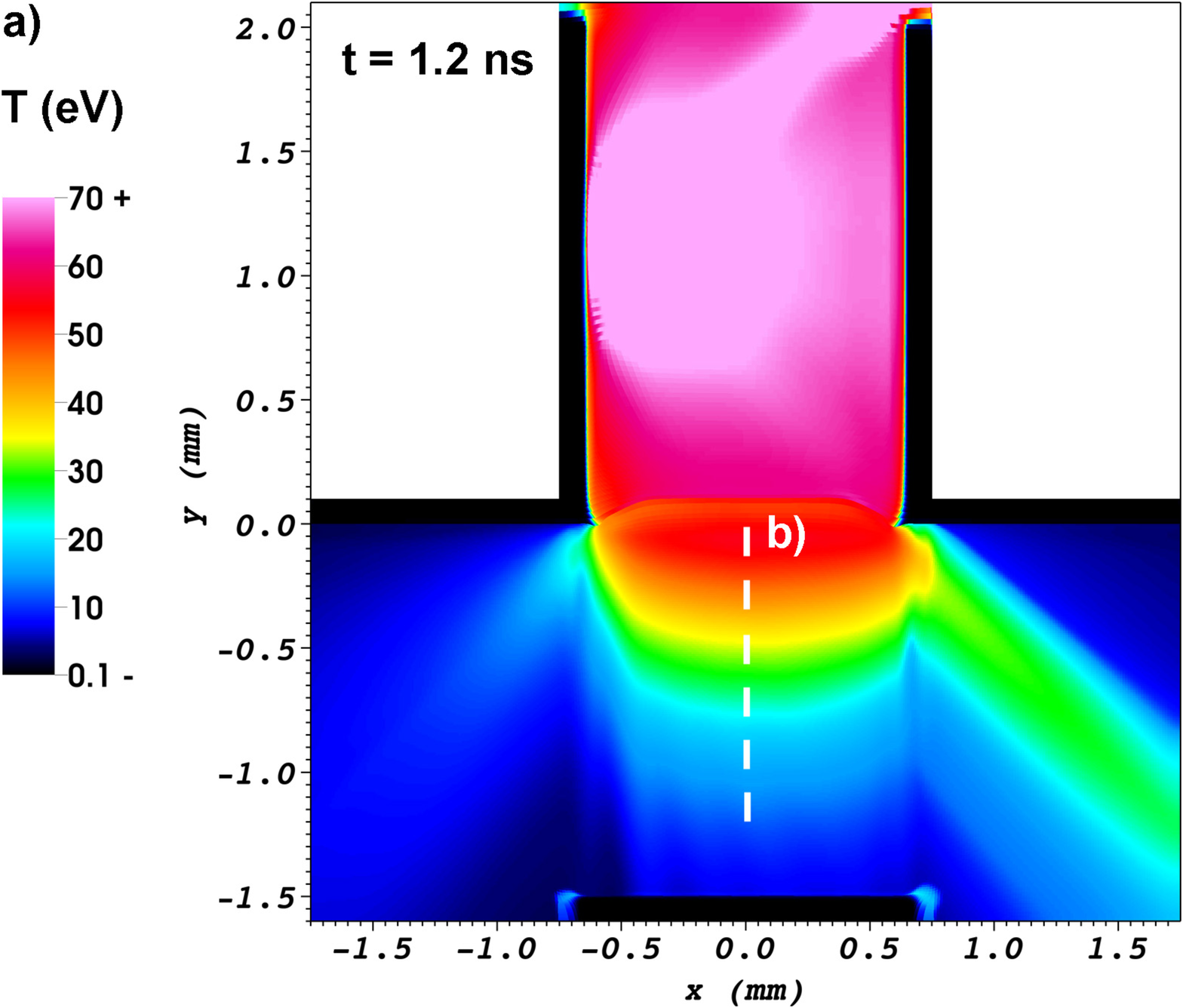}\\[0.4cm]
\includegraphics*[width=0.9\columnwidth]{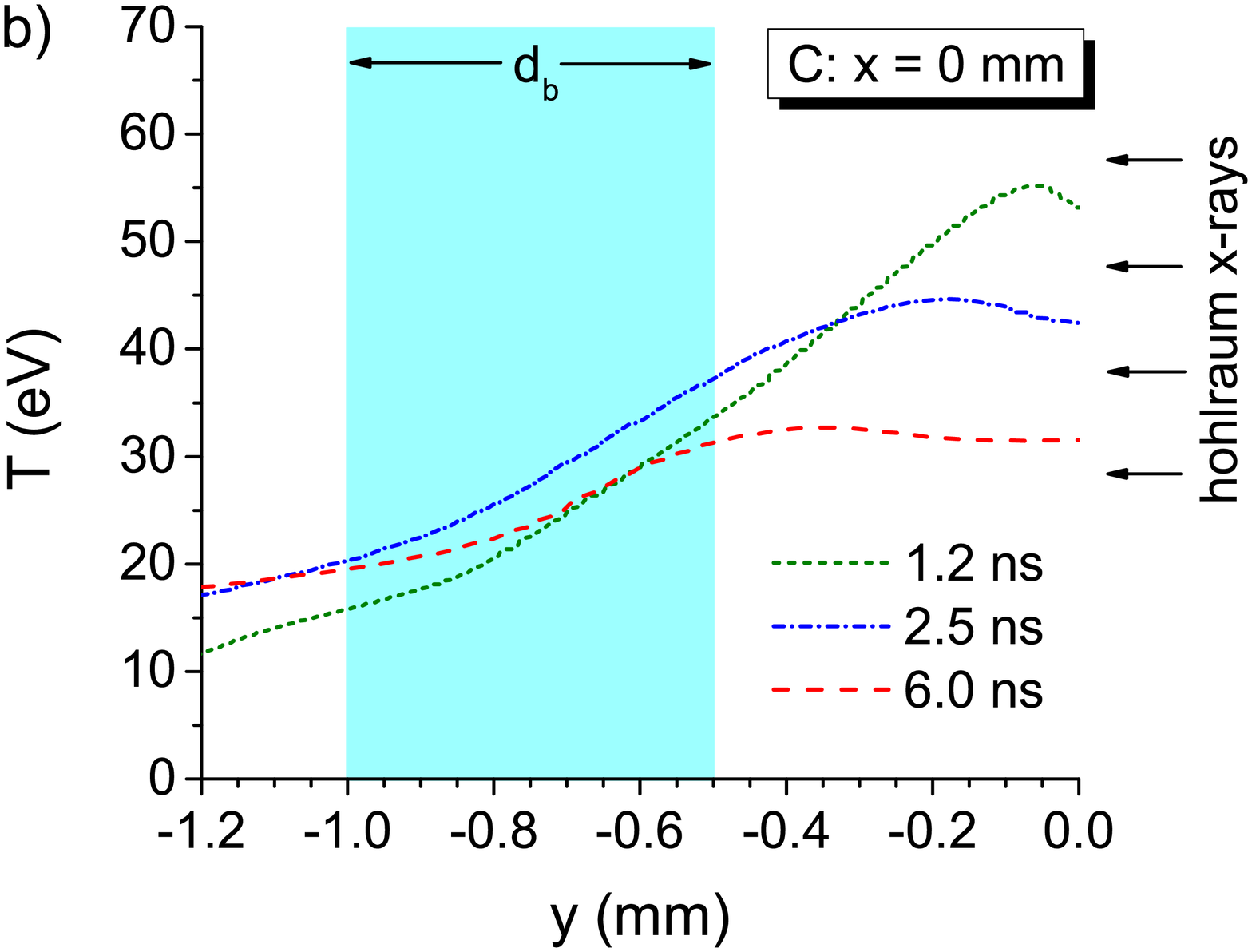}
\caption{\label{f:foam-T} (color online) a) Color contour plot of the matter temperature $T$ at $t=1.2$~ns, and b)~vertical temperature profiles through the carbon block center at three selected times, $t=1.2$, 2.5, and 6~ns; the shaded (cyan) vertical strip of width $d_b$ marks the ion beam aperture.}
\end{figure}

As may be expected, most of the energetic X-rays, originating from the focal laser spot plasma, enter and heat the carbon block already during the laser pulse. The distribution of the matter temperature by the end of the laser pulse at $t=1.2$~ns is shown in Fig. \ref{f:foam-T}a. At this time a strong spatial variation of the carbon plasma temperature, ranging from values below 10~eV to those above 55~eV, can be observed. Furthermore, some X-rays, mainly those which originate from the laser spot at the left hohlraum wall, pass through the carbon block at its upper corners and ``shine'' into the low-density carbon vapor fill.

\begin{figure}
\centering
\includegraphics*[width=0.9\columnwidth]{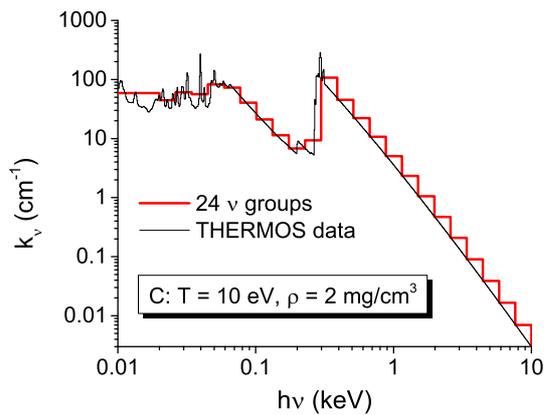}
\caption{\label{f:knu_C} Spectral absorption coefficient $k_{\nu}$ of carbon at $T=10$~eV and $\rho=2$~mg~cm$^{-3}$ used in the simulations: the original \mbox{THERMOS} code data (thin solid curve) are shown together with the group-averaged values for 24 (thick solid curve) selected spectral groups.}
\end{figure}

The vertical temperature gradient across the plasma and its relaxation in time are illustrated by three selected temperature profiles through the center of the carbon block in Fig.~\ref{f:foam-T}b. For a proper qualitative understanding of the carbon heating process one has to consider the spectral absorption coefficient $k_{\nu}$ of carbon, shown in Fig.~\ref{f:knu_C} for $T=10$~eV and $\rho =2$~mg~cm$^{-3}$. By comparing the plots in Figs.~\ref{f:hr-spectrum} and \ref{f:knu_C} one sees that the first spectral peak of strong emission from the hot ($T \simeq 300$~eV) gold plasma in the laser spot practically coincides with the transparency window $h\nu = 0.1$--0.3~keV of carbon just below its $K$-edge, where $k_{\nu} \approx1$~mm$^{-1}$; the photons with $h\nu \gtrsim 0.8$~keV from the second spectral emission peak also have $k_{\nu} \lesssim 1$~mm$^{-1}$. In other words, for a large portion of the hohlraum radiation emitted during the laser pulse the carbon foam has an optical thickness of $\approx 1$. The latter means that the carbon foam is practically instantaneously (i.e.\ synchronously with the temporal laser power profile) heated by a flash of X-rays from the laser spot over the entire foam volume to an average temperature of $T \approx 30$~eV, varying by about a factor 4 across a distance of 1~mm. At the same time, about 20\% of the hohlraum X-ray emission, generated during the laser pulse, passes through the foam and directly hits the copper support plate.

Note that the described heating dynamics of our low-density carbon foam is in many respects similar to that diagnosed by Gregori {\em et al.} \cite{GrGl.08} in earlier experiments at the OMEGA facility --- though the latter used a 100 times denser foam and about 50 times more powerful laser pulses. In both cases the foam is heated in a clearly supersonic regime by quasi-thermal X-rays, whose effective radiation temperature significantly exceeds the peak electron temperature of the heated sample.

\subsection{Temperature equilibration phase}

\begin{figure*}
\centering
\includegraphics*[width=0.93\textwidth]{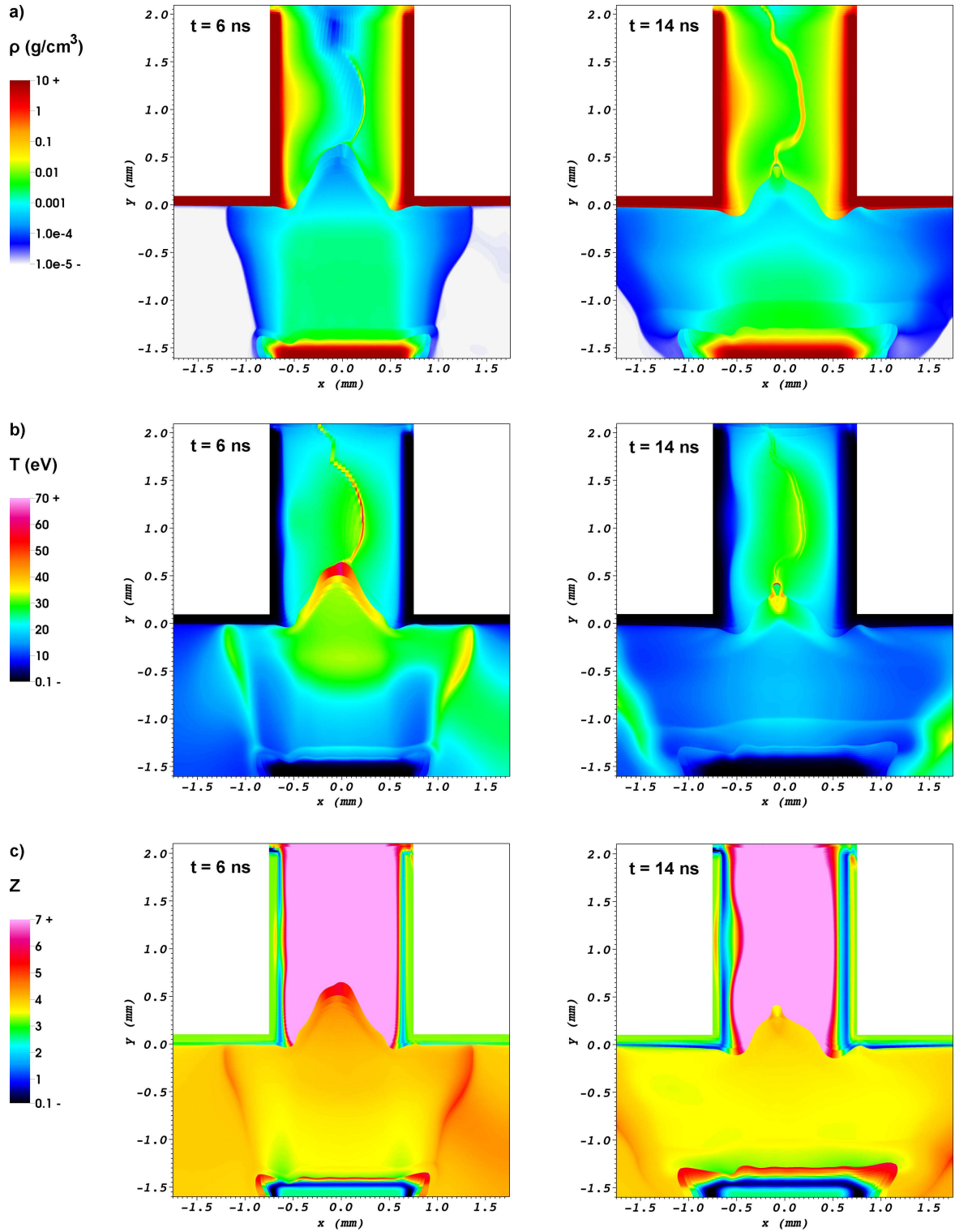}
\caption{\label{f:foam-VisIt} (color online) Color contour plots of a)~the matter density $\rho$, b)~the matter temperature $T$, and c)~the LTE ionization degree $Z$ of the whole experimental configuration at $t=6$ and $14$~ns. In the middle of the hohlraum one observes a filamentary structure, evolving from the collision of the ablated gold plasma flows. Material interfaces can be identified as sharp discontinuities on the $Z$ plot. Inside the carbon block one can discern an almost planar shock front, propagating upwards from the copper support plate and reaching $y=-1$~mm at $t=13.5$~ns.}
\end{figure*}

After the laser is turned off, the hohlraum continues to glow in soft X-rays on a time scale of tens of nanoseconds, letting out the energy accumulated in its wall material during the laser pulse. Figure \ref{f:foam-VisIt} shows the 2D distributions of the density $\rho$, the matter temperature $T$, and the LTE ionization degree $Z$ for the whole experimental configuration at $t=6$ and 14~ns. From these plots some important observations can be made for the ion-stopping measurements.

Firstly, one can clearly see that the expanding carbon plasma pushes back the gold plasma from the hohlraum, which means that one should not fear the ablated gold to get into the way of the ion beam. Secondly, as a significant energy portion of the main X-ray flash, generated by the laser pulse, penetrates through the translucent carbon foam and is absorbed by the copper holder, the resulting pressure discontinuity between copper and foam launches a shock wave into the foam. This shock front propagates towards the hohlraum and enters the ion-beam aperture $-1.0$~mm~$<y<-0.5$~mm at $t\approx13.5$~ns, i.e.\ late enough to perform the measurements at earlier times. Finally, the simulation shows that a reasonably homogeneous plasma volume with $Z \approx 3.7$--3.8 and $T\approx 23$--27~eV is created along the ion beam path, quite suitable for the ion-stopping measurements within a certain limited time window.

\begin{figure}
\centering
\includegraphics*[width=0.9\columnwidth]{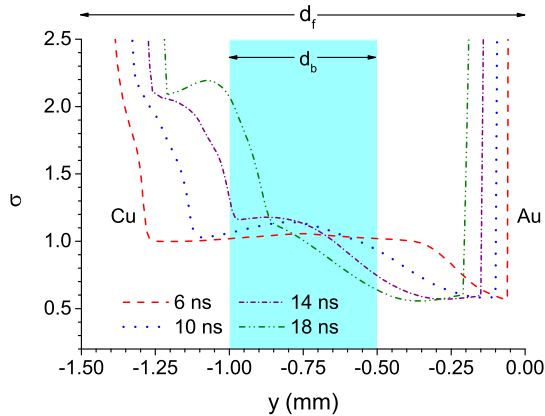}
\caption{\label{f:column-rho} Normalized column density $\sigma(t,y)$ along the ion trajectories parallel to the x-axis (see Figs. \ref{f:3D-setup} and \ref{f:2D-setup}) as a function of the $y$-coordinate for four selected times $t=6$, 10, 14, and 18~ns. The positions of the copper and gold plasma fronts are easily identified as almost vertical walls on the left and on the right. The shock front (smeared due to 2D effects) in the foam, pushed by the expanding copper, reaches the ion beam aperture $d_b$ at $t=13.5$~ns.}
\end{figure}

\begin{figure}
\centering
\includegraphics*[width=0.9\columnwidth]{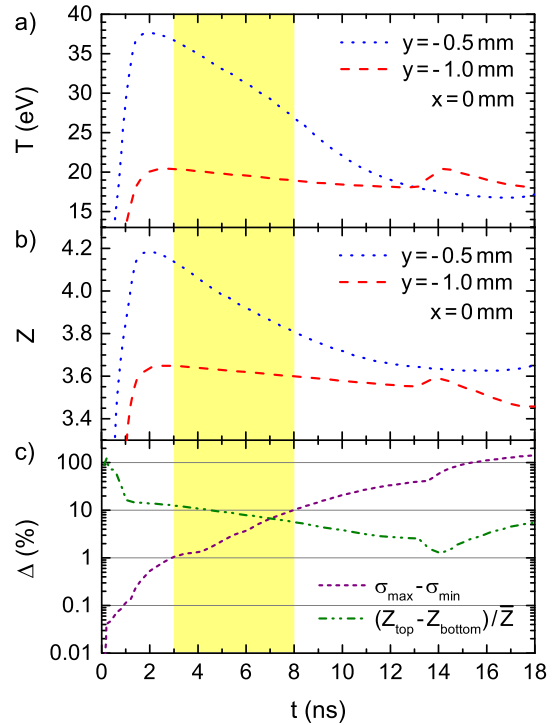}
\caption{\label{f:foam-T-Z-his} Temporal evolution of a) the matter temperature $T$ and b) the LTE ionization degree $Z$ at the top ($y=-0.5$~mm) and the bottom ($y=-1.0$~mm) of the ion beam cross-section in the center of the foam block ($x=0$). Figure c) shows the maximum spatial variation of the normalized column densities $\sigma(t,y)$ within the ion beam cross-section and the difference between the top and bottom ionization degree values weighted by their mean value $\bar{Z}=\frac{1}{2}(Z_{top}+Z_{bottom})$ as functions of time. A propitious time window for the ion-stopping measurements is marked as a shaded (yellow) vertical strip.}
\end{figure}

The appropriate time window $[t_1,t_2]$ for the ion-stopping measurements can be evaluated by inspecting the data presented in Figs.~\ref{f:column-rho} and \ref{f:foam-T-Z-his}. Figure~\ref{f:column-rho} shows the normalized column density
\begin{equation}\label{sig_x=}
    \sigma(t,y) = \left[\int\rho(t,x,y)\,dx \right] \left[\int\rho(0,x,y)\,dx \right]^{-1}
\end{equation}
along the supposed ion trajectories parallel to the x-axis (see Figs. \ref{f:3D-setup} and \ref{f:2D-setup}) as a function of the $y$-coordinate at four selected times $t=6$, 10, 14, and 18~ns. For high-quality ion stopping measurements, it is important to have the values of $\sigma(t,y)$ as close to 1 as possible for the entire ion pulse duration $t_b \approx 3$~ns over the entire beam aperture $-1.0$~mm~$<y<-0.5$~mm. The eventual significant departures of $\sigma(t,y)$ from 1 are caused by three effects: (i)~the 2D lateral expansion of the carbon plasma beyond the confining walls of the copper holder, (ii)~compression of the foam material by the shock front propagating from the bottom of the carbon block, and (iii)~swelling of the gold walls near the hohlraum edges. Our results indicate that for $t \lesssim 8$~ns the maximum deviations of $\sigma(t,y)$ from unity within the diameter of the ion beam $d_b$ do not exceed 10\%. The $t=14$ and 18~ns profiles in Fig.~\ref{f:column-rho} are already significantly perturbed by the effects (i) and (ii).

Figures \ref{f:foam-T-Z-his} a) and b) show the temporal evolution of the matter temperature $T$ and the ionization degree $Z$ at two points $(x,y)=(0,-0.5)$ and $(0,-1.0)$ that are of special interest from the point of view of the ion-stopping measurements. Figure \ref{f:foam-T-Z-his} c) shows the difference between both ionization degree values and the maximum spatial variation of $\sigma(t,y)$ within the ion beam aperture. One sees that the temperature and ionization become quite uniform for $t \gtrsim 10$~ns --- which would be a good time for the ion bunch to arrive if not already significant perturbations to the $\sigma(t,y)$ values occurred.

Finally, as a result of complex interplay between the above discussed physical effects, the best compromise for the ion measurements in the analyzed configuration should roughly fall into the range 3~ns${} \lesssim t \lesssim 8$~ns --- as is indicated in Fig.~\ref{f:foam-T-Z-his} with a shaded vertical strip. Note that the upper boundary of this window can easily be expanded by increasing the size of the foam block: every extra 100~${}\mu$m of the foam size would add roughly 1~ns to the favorable time span.

\section{Conclusion \label{s:concl}}

The results of a 2D radiation-hydrodynamic study of a combined hohlraum-foam target used to perform ion-stopping measurements in a quasi-homogeneously ionized plasma are presented. Our principal goal was to investigate how uniform and for how long a sample plasma column can be created in this type of laser-driven targets. We have analyzed the principal physical processes, which determine the spatial structure and dynamics of the sample plasma, and demonstrated how their combined effect defines an optimum --- to ensure clean ion-stopping measurements in a plasma --- time window for accepting a $\simeq 3$-ns long bunch of fast ions. For the target parameters used in this work, this time interval turns out to be between $3$~ns and $8$~ns after the onset of the laser pulse. We find that within this window the time and space variations of such key parameters as the column mass density along the ion trajectories and the plasma ionization degree do not exceed $\pm 7\%$. On the basis of our analysis one can readily identify the guidelines for target modifications that could improve the quality of the ion-stopping measurements in this type of laser-plasma targets.

At the same time we have to admit serious limitations of the present 2D simulations in what concerns direct comparison with the experimental data. First of all, it can hardly be expected that the simulated 2D configuration sufficiently accurately represents the essentially 3D experimental arrangement: the qualitative parameters of the carbon plasma uniformity may be significantly modified by the 3D effects in the hydrodynamic motion and the radiation transport. Secondly, the experimentally used cellulose-triacetate (C$_{12}$H$_{16}$O$_8$) foam contains a significant fraction of oxygen atoms, which noticeably modify the spectral absorption coefficient around the K-edges of carbon and oxygen. Hence, more accurate opacity data for the foam material are needed before a detailed comparison with the experiment is to be made.

Thirdly, some uncertainty arises from the fact that the radiation transport was simulated with the LTE spectral absorption and emission coefficients. Although the LTE approximation appears to be well justified for times after the end of the laser pulse, the non-LTE effects in the laser-ablated gold plasma with $T=300$--400~eV during the laser pulse may noticeably alter the X-ray spectrum emerging from the hohlraum at $t< 1.2$~ns. Finally, since we used a simplified model for the laser light transport without refraction and reflection, no estimate can be given for possible effects of the reflected (off the hohlraum wall) laser light hitting directly the carbon foam and other parts of the hohlraum. To study such effects, a more advanced laser propagation package is needed than presently available in the RALEF code.

\section*{Acknowledgments \label{s:acknow}}
This work was supported by the ExtreMe Matter Institute EMMI in the framework of the Helmholtz Alliance Program HA216/EMMI, by the Bundesministerium f\"ur Bildung und Forschung BMBF (Project 05P12RFFTR), by the Helmholtz International Center for FAIR (HIC for FAIR), and by the J\"ulich Supercomputing Centre JSC.

\appendix

\section{\label{s:2D3Dconv} Reduction of a 3D hohlraum to a 2D configuration}

Reduction of an intrinsically 3D problem to two dimensions requires an additional spatial symmetry to be imposed on the original 3D configuration, which may be either a translational invariance (along the $z$-axis) or an invariance with respect to rotation around a fixed $z$-axis. In result one would want the new 2D configuration to reproduce as closely as possible all the main dynamic features of the original 3D problem. In our case it means that the simulated 2D configuration must represent a certain 2D cut of the original 3D hohlraum with all the key dimensions taken from the latter. Having preserved the temporal and spatial shapes of the driving laser pulse, we are left with only one undetermined parameter for the 2D case, namely, the total input laser energy, which we will denote as $\tilde{E_l}$ for the 2D case, and as $E_l$ for the original 3D case. Note that the original value $E_l$ must be recalculated (rescaled) to the ``equivalent'' 2D value $\tilde{E_l}$ already because in the case of translational invariance these two quantities have different physical dimensions.

Since in hohlraum-type targets both the matter motion and the intensity (as well as the spectrum) of the hohlraum radiation are controlled by the incident radiation-energy fluxes per unit surface area of the hohlraum interior, we assume that the ``physically equivalent'' 2D hohlraum must {\em on average} accept the same amount of energy per unit inner surface area as the original 3D hohlraum. Mathematically this condition can be expressed as the following two equations of the global energy balance in a hohlraum
\begin{eqnarray}\label{3D2D:balance1}
  E_l&=&F_wS_w+F_hS_h,
  \\ \label{3D2D:balance2}
  \tilde{E_l}&=&F_w\tilde{S}_w+F_h\tilde{S}_h.
\end{eqnarray}
Here $S_w$ and $S_h$ are, respectively, the total surface areas of the inner 3D hohlraum wall and of all its holes, $\tilde{S}_w$ and $\tilde{S}_h$ are the corresponding quantities in the 2D hohlraum, $F_w$ is the radiation-energy fluence (measured in J/cm${}^2$) absorbed by the inner hohlraum walls, and $F_h$ is the radiation-energy fluence which escapes the hohlraum through its holes. If the hohlraum radiation is characterized by the equivalent black-body temperature $T_r(t)$, then $F_h$ can be calculated as
\begin{equation}
  F_h = \sigma_{SB} \int T_r^4\,dt,
\end{equation}
where $\sigma_{SB}$ is the Stefan-Boltzmann constant. As a word of caution it should be noted that, because it is only the time- and space-averaged quantities $F_w$ and $F_h$ that are ensured to have the same values in the original 3D hohlraum and its 2D counterpart, one cannot expect that all the details of the 3D problem should be adequately reproduced by a corresponding 2D simulation.

Since the surface areas $S_w$, $S_h$, $\tilde{S}_w$, and $\tilde{S}_h$ are all known, Eqs. (\ref{3D2D:balance1}) and (\ref{3D2D:balance2}) yield the following expression for the rescaled input energy
\begin{equation}\label{3D2D:Escal}
  \tilde{E}_l=E_l\frac{\tilde{S}_w+q_{hw}\tilde{S}_h}{S_w+q_{hw}S_h},
\end{equation}
where
\begin{equation}
  q_{hw}=\frac{F_h}{F_w}
\end{equation}
is an unknown dimensionless factor. It can be found by the method of successive approximations as follows. In the zeroth approximation, one can set $F_h=F_w$, i.e. assume that $q_{hw}=q^{(0)}_{hw}=1$, and perform a 2D simulation with
\begin{equation}
  \tilde{E}_l=\tilde{E}_l^{(0)}=E_l\frac{\tilde{S}_w+\tilde{S}_h}{S_w+S_h}.
\end{equation}
Making use of the results of this simulation, one can evaluate the first-order value of $q_{hw}$ as
\begin{equation}\label{3D2D:qfirstorder}
  q_{hw}^{(1)}=\frac{\tilde{S}_w}{\tilde{S}_h}\frac{\tilde{E}_{r,out}^{(0)}}{
  \tilde{E}_l^{(0)}-\tilde{E}_{r,out}^{(0)}},
\end{equation}
where $\tilde{E}^{(0)}_{r,out}$ is the total amount of radiative energy, which escapes the hohlraum in the zero-order 2D run. Equation~(\ref{3D2D:qfirstorder}) is a direct consequence of the global energy balance relation (\ref{3D2D:balance2}) combined with
\begin{equation}
  \tilde{E}_{r,out}=F_h\tilde{S}_h.
\end{equation}
Having substituted  $q^{(1)}_{hw}$ into Eq. (\ref{3D2D:Escal}), one obtains a first-order estimate $\tilde{E}_l^{(1)}$ for the rescaled input energy $\tilde{E}_l$. Higher-order approximations are usually not needed. Note that hohlraums with walls of a heavy metal typically have $q_{hw} \approx 2$--3.

%\section*{References}
\bibliographystyle{model1-num-names}
\bibliography{Foam2012}

\end{document}